# Epitaxial inversion of spontaneous polarization in ε-$Ga_2O_3$

**Yan Wang[1,2*], Zhigao Xie[1,2], Weihua Tang[1*], Chee Keong Tan[1,2,3,4,5*]**

[1]*College of Integrated Circuit Science and Engineering, Nanjing University of Posts and Telecommunications, Nanjing 210023, China*

[2]*Advanced Materials Thrust, Function Hub, The Hong Kong University of Science and Technology (Guangzhou), Guangzhou 511453, Guangdong, China*

[3]*Department of Electronic and Computer Engineering, The Hong Kong University of Science and Technology, Hong Kong, China*

[4]*Guangzhou Municipal Key Laboratory of Materials Informatics, The Hong Kong University of Science and Technology (Guangzhou), Guangzhou 511453, Guangdong, China*

[5]*Guangzhou Municipal Key Laboratory of Integrated Circuits Design, The Hong Kong University of Science and Technology (Guangzhou), Guangzhou 511453, Guangdong, China*

* Corresponding author.

Email addresses: ywang950@connect.hkust-gz.edu.cn (Y. Wang); whtang@njupt.edu.cn (W. Tang); 20230193@njupt.edu.cn & cheekeongtan@hkust-gz.edu.cn (C. K. Tan)

## ABSTRACT

Polar wide bandgap semiconductors offer the unique capability to manipulate internal electric fields and induce two-dimensional electron gases (2DEG). The emerging orthorhombic ε-$Ga_2O_3$ is a promising candidate owing to its large spontaneous polarization ($P_{sp}$). However, exploiting this material is hindered by fundamental ambiguities surrounding its absolute $P_{sp}$ vectors and the inability to govern its epitaxial direction. Here we show the resolution of the absolute $P_{sp}$ vectors in ε-$Ga_2O_3$ and control of its macroscopic polarity via substrate engineering. By correlating interferometric piezoresponse with atomic configurations, we establish an identification criterion where opposing polarities are assigned through the geometric elevation within an asymmetric four-atom sequence. Guided by this signature, we demonstrate that epitaxy on Al-polar AlN and sapphire yields uniformly Ga-polar (upward $P_{sp}$) and O-polar (downward $P_{sp}$) architectures, respectively, whereas deteriorated crystallinity disrupts this registry and triggers mixed-polarity. This atomic-to-macroscopic correlation eliminates long-standing ambiguities regarding absolute polar orientations in non-centrosymmetric oxides. Analogous to mature III-Nitride architectures, this blueprint provides the foundational platform for designing macroscopic polarization discontinuities at heterointerfaces, unlocking ε-$Ga_2O_3$ for advanced polarization electronics.

## 1. Introduction

Ultra-wide bandgap (UWBG) gallium oxide ($Ga_2O_3$) has emerged as a premier semiconductor platform, driving a paradigm shift in the development of next-generation power electronics and high-frequency radio-frequency devices [1-5]. While extensive research has centered on the thermodynamically stable monoclinic β-phase, the non-centrosymmetric orthorhombic polymorph, ε-$Ga_2O_3$ (strictly identical to κ-$Ga_2O_3$, space group $Pna2_1$), offers fundamentally distinct physical capabilities. Specifically, the intrinsic breaking of inversion symmetry in the ε-phase lattice generates a robust spontaneous polarization ($P_{sp}$) of 23–24 μC/cm$^2$ along its c-axis [6-10]. This inherent polar nature elevates ε-$Ga_2O_3$ beyond a conventional wide-bandgap semiconductor, establishing it as a highly tunable polar-oxide matrix for advanced nanoscale polarization engineering.

The ability to engineer this structural polarity is the cornerstone of advanced semiconductor device physics. In mature III-Nitride systems, the strategic design of macroscopic polarization discontinuities serves as the fundamental mechanism for inducing high-density, dopant-free two-dimensional electron gases (2DEG) at heterointerfaces [11-13]. By precisely manipulating these polarization vectors, further breakthroughs have been achieved, such as efficient polarization-induced hole doping and the realization of two-dimensional hole gases (2DHG) [14, 15]. Advancing beyond interfacial discontinuities, the rigorous macroscopic control over absolute crystal orientation (e.g., Ga-polar versus N-polar GaN) enables the profound manipulation of internal electrostatic boundary conditions [16-18]. A quintessential demonstration of this advanced polar mastery is the recent development of monolithic dualtronics by van Deurzen et al., where N-polar and Ga-polar surfaces of a single wafer are selectively exploited to integrate advanced high electron mobility transistors (HEMTs) and light-emitting diodes (LEDs) [19]. Translating this comprehensive polarity-driven paradigm to UWBG oxides represents the trajectory for unlocking the full potential of ε-$Ga_2O_3$-based HEMTs.

However, accurately identifying and controlling the absolute polarization direction in emergent crystal systems is treacherous. The profound difficulty of this metrological task is historically underscored by the early research on III-Nitrides, where the scientific community engaged in years of intense debate, misassigning the absolute magnitude and even the direction of the spontaneous polarization in AlN and GaN before advanced first-principles theories and sophisticated metrology corrected the consensus [20-22]. Today, the study of ε-$Ga_2O_3$ faces an identical historical bottleneck. Conventional piezoelectric force microscopy (PFM) relies heavily on relative contrast and arbitrary phase offsets, heavily convoluted by surface topography. Consequently, establishing the true, absolute $P_{sp}$ direction of ε-$Ga_2O_3$ independent of external references remains unachieved. The fundamental physical relationship between the macroscopic polar response, the atomic-scale geometric configurations, and the substrate-induced epitaxial registry remains obscured.

In this work, we overcome these long-standing metrological and physical ambiguities by resolving the absolute $P_{sp}$ vectors in ε-$Ga_2O_3$ and demonstrating control over its macroscopic polarization via substrate engineering. Deploying quadrature phase differential interferometry (QPDI) PFM, we directly quantify the absolute out-of-plane lattice displacement. We correlate this macroscopic absolute phase with the atomic configurations resolved by scanning transmission electron microscopy (STEM), revealing that the absolute polarity is strictly dictated by the geometric elevation or depression within a characteristic four-atom structural sequence. This structural-polar correlation is rigorously validated by first-principles Berry-phase calculations. Through this framework, we establish the fundamental atomistic rules for polarization control by demonstrating that growth on Al-polar AlN/sapphire substrates allows for the intentional synthesis of Ga-polar domains with an upward-pointing $P_{sp}$, whereas growth on sapphire templates deterministically results in O-polar films with a downward-pointing $P_{sp}$. However, our experimental macroscopic mapping further reveals that while pristine epitaxy preserves this substrate-induced registry, crystal quality degradation effectively screen the substrate's electrostatic boundary conditions, triggering a physical relaxation into mixed-polarity domains. Coupled with extreme high-voltage testing that confirms these orientations as thermodynamically pinned, non-ferroelectric states, this framework provides the epitaxial blueprint required for future UWBG polarization engineering.

## 2. Results and Discussion

### 2.1 Crystal quality and epitaxial relationship of ε-$Ga_2O_3$

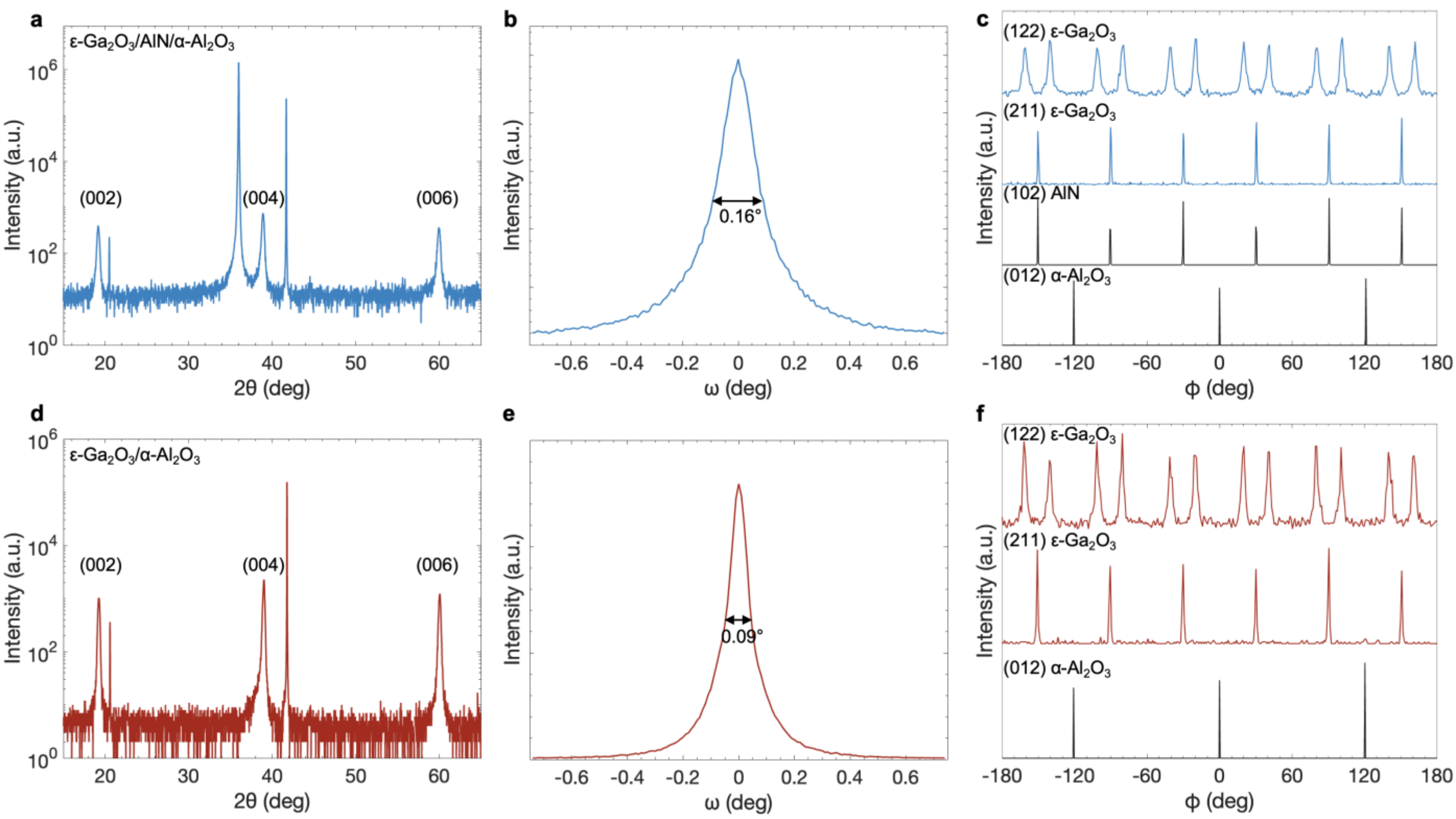

**Fig. 1 | Structural characterization of ε-$Ga_2O_3$ epitaxial films.** (**a** to **c**) XRD profiles of the ε-$Ga_2O_3$/AlN/α-$Al_2O_3$ heterostructure, exhibiting (**a**) the 2θ-scan, (**b**) the ω-scan rocking curve of the ε-$Ga_2O_3$ (004) reflection (FWHM = 0.16°), and (**c**) the ϕ-scans of the ε-$Ga_2O_3$ (122) and (211), AlN (102), and α-$Al_2O_3$ (012) planes. (**d** to **f**) Corresponding profiles of the ε-$Ga_2O_3$ film grown directly on the α-$Al_2O_3$ substrate, showing (**d**) the 2θ-scan, (**e**) the ω-scan rocking curve of the ε-$Ga_2O_3$ (004) reflection (FWHM = 0.09°), and (**f**) the ϕ-scans of the ε-$Ga_2O_3$ (122) and (211) and α-$Al_2O_3$ (012) planes.

The crystalline quality and phase purity of the heteroepitaxial structures were systematically evaluated using high-resolution X-ray diffraction (XRD). As presented in the 2θ profiles (Fig. 1a and 1d), only the (00*l*) reflections of ε-$Ga_2O_3$ are detected alongside the substrate and template peaks. Specifically, for the ε-$Ga_2O_3$/AlN/α-$Al_2O_3$ heterostructure, the ε-$Ga_2O_3$ (002), (004), and (006) diffraction peaks emerge precisely at 2θ = 19.23°, 38.89°, and 59.99°, respectively, accompanied by the underlying AlN template peak located at 35.99°. For the ε-$Ga_2O_3$ film grown directly on the α-$Al_2O_3$ substrate, the analogous ε-$Ga_2O_3$ (002), (004), and (006) peaks are positioned at 19.23°, 39.05°, and 60.09°. The exclusive presence of these specific reflections confirms a strict c-axis out-of-plane orientation and absence of secondary polymorphs. The crystal quality was quantitatively assessed via ω-scan rocking curves of the ε-$Ga_2O_3$ (004) plane. The extracted full width at half maximum (FWHM) values are 0.16° for the ε-$Ga_2O_3$/AlN/α-$Al_2O_3$ heterostructure (Fig. 1b) and 0.09° for the ε-$Ga_2O_3$ grown directly on α-$Al_2O_3$ (Fig. 1e), indicating exceptional out-of-plane crystalline perfection in both configurations. Furthermore, ϕ-scans were executed to determine the in-plane epitaxial relationships and domain architectures (Fig. 1, c and f). In both configurations, the ε-$Ga_2O_3$ (211) reflections exhibit six distinct peaks. Notably, these (211) peaks are perfectly aligned with the AlN (102) reflections in the template-assisted sample, while exhibiting a strict 30° shift relative to the α-$Al_2O_3$ (012) substrate peaks. Concurrently, the ε-$Ga_2O_3$ (122) reflections display a 12-fold symmetry, with peaks symmetrically distributed at ±10° relative to the respective (211) peak positions. This identical azimuthal diffraction signature across both samples rigorously validates the formation of the same rotational domains regardless of the underlying interface. This crystallographic registry establishes the epitaxial relationship where <100> ε-$Ga_2O_3$ aligns with <11-20> sapphire and <010> ε-$Ga_2O_3$ matches

<10-10> sapphire, consistent with established heteroepitaxial growth mechanisms reported for similar oxide semiconductor systems [23-25].

## 2.2 Atomic-scale structural inversion

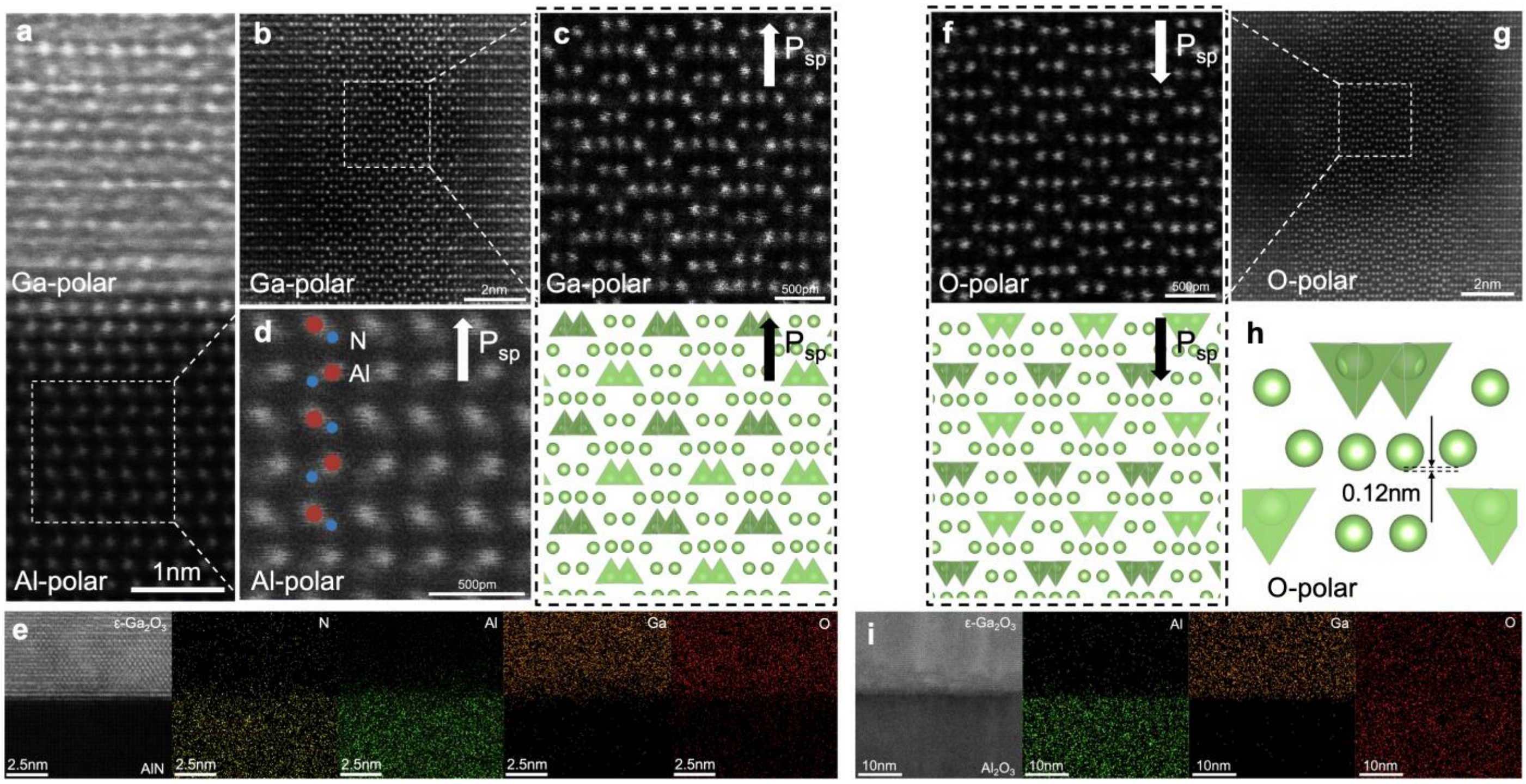

**Fig. 2 | Atomic-scale structural inversion and polarity identification in ε-$Ga_2O_3$ heterostructures.** (**a** to **e**) Characterization of the Ga-polar ε-$Ga_2O_3$/AlN/α-$Al_2O_3$ heterostructure. (**a**) STEM image of the interface between the ε-$Ga_2O_3$ film and the Al-polar AlN template. (**b**) STEM image of the ε-$Ga_2O_3$ lattice. (**c**) Magnified view of the selected region in (**b**) matched with the corresponding projected atomic model. The tetrahedral sites are highlighted using polyhedra, exhibiting an upward orientation of the tetrahedral vertices. (**d**) Magnified view of the AlN lattice showing the upward-pointing atomic arrangement of Al and N columns, corresponding to its intrinsic Al-polar nature with an established upward $P_{sp}$. (**e**) Cross-sectional STEM-EDS elemental mappings of N, Al, Ga, and O. (**f** to **i**) Characterization of the O-polar ε-$Ga_2O_3$ film grown directly on the α-$Al_2O_3$ substrate. (**f**) High-resolution STEM image of the O-polar ε-$Ga_2O_3$ lattice. (**g**) Magnified view of the region highlighted in (**f**) aligned with the corresponding atomic model. The highlighted tetrahedral polyhedra exhibit downward-pointing vertices, revealing a complete structural inversion relative to the configuration in (**c**). (**h**) Schematic atomic configuration detailing the 0.12 nm atomic displacement characteristic of the O-polar structure. (**i**) Cross-sectional STEM-EDS elemental mappings of Al, Ga, and O at the interface. *(Note: While the upward $P_{sp}$ of the Al-polar AlN template is a known intrinsic property, the absolute $P_{sp}$ directions of the ε-$Ga_2O_3$ layers cannot be solely determined from these geometric observations. The structural inversion observed herein merely confirms a relative anti-parallel polarity between the two configurations. The precise $P_{sp}$ vectors indicated by the arrows are rigorously established by subsequent PFM measurements and density functional theory simulations.)*

To differentiate the microstructural architectures of the two distinct epitaxial systems, the film grown on the Al-polar AlN template is strictly designated as Ga-polar ε-$Ga_2O_3$, whereas the film deposited directly onto the bare α-$Al_2O_3$ substrate is defined as O-polar ε-$Ga_2O_3$. Cross-sectional STEM image provides highly sensitive atomic imaging, was employed to dissect their atomic-scale interfacial configurations and local lattice symmetries. Fig. 2a displays the sharp interface between the Ga-polar ε-$Ga_2O_3$ and the Al-polar AlN template. The magnified view in Fig. 2d explicitly resolves the alternating arrangement of Al and N atomic columns. Notably, the Al atoms (red spheres) are positioned directly above the N atoms (blue spheres), a well-established atomic configuration that inherently dictates an upward $P_{sp}$ direction, consistent with prior experimental and theoretical validations [20, 26]. Fig. 2b presents the atomic arrangement of the Ga-polar ε-$Ga_2O_3$ film, where the rotational domains are clearly visible (Fig. S1). Fig. 2c is a magnified view of Fig. 2b, specifically capturing the (100) plane of the ε-$Ga_2O_3$ lattice, as this specific projection is uniquely required for determining the crystal orientation based on atomic arrangements. In this view, a characteristic sequence of four consecutive atoms is identified directly above the highlighted tetrahedral sites. These four atoms are definitively not on a single horizontal plane; rather, the central two atoms are situated slightly higher than the adjacent outer two. Coupled with this specific atomic arrangement, the vertices of the corresponding tetrahedral sites point strictly upward.

Conversely, Fig. 2g exhibits the STEM image of the O-polar ε-$Ga_2O_3$ film, where the rotational domains are similarly well-resolved, corroborating the preceding XRD analysis (Fig. S2). The magnified view of the O-polar (100) plane (Fig. 2f) reveals a stark structural contrast: the characteristic four-atom sequence is now located below the highlighted tetrahedral sites. Furthermore, the central two atoms in this sequence are positioned slightly lower than the outer two, demonstrating a complete reversal of the atomic orientation compared to the Ga-polar architecture. As quantified by the corresponding unit cell model (Fig. 2h), this fundamental positional displacement between the atomic arrangements is 0.12 nm. Finally, the compositional abruptness and elemental integrity of both epitaxial systems are confirmed by the cross-sectional STEM-EDS elemental mappings (Fig. 2e and Fig. 2i), which exhibit distinct and interdiffusion-free boundaries for Al, Ga, N, and O across their respective interfaces.

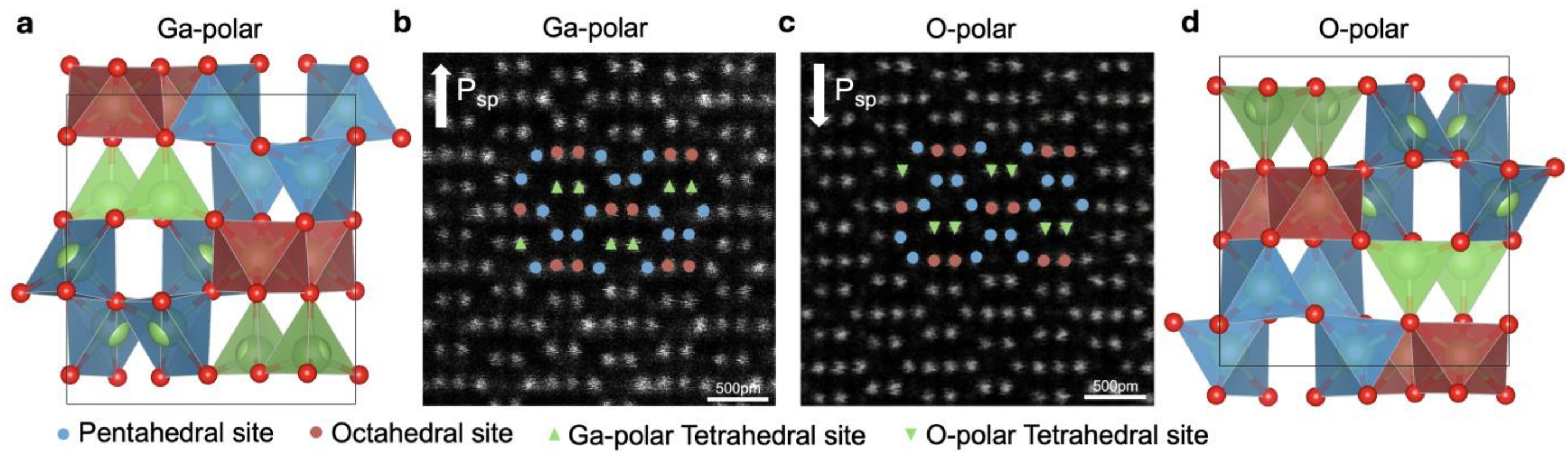

**Fig. 3 | Atomic-site mapping and polyhedral configurations of the polarity-inverted ε-$Ga_2O_3$ lattices.** (**a**) Projected atomic model of the Ga-polar ε-$Ga_2O_3$ unit cell, delineating the distinct gallium coordination environments: pentahedral (blue), octahedral (red), and tetrahedral (green) polyhedra. (**b**) Annotated high-resolution STEM image of the Ga-polar ε-$Ga_2O_3$ film. The overlaid markers identify the projected Ga atomic columns corresponding to the pentahedral (blue circles), octahedral (red circles), and tetrahedral (green upward triangles) sites, consistent with the model in (**a**). (**c**) Annotated high-resolution STEM image of the O-polar ε-$Ga_2O_3$ film. The overlaid markers map the corresponding pentahedral (blue circles), octahedral (red circles), and structurally inverted tetrahedral sites (green downward triangles). (**d**) Projected atomic model of the O-polar ε-$Ga_2O_3$ unit cell, illustrating the geometric inversion of the polyhedral network relative to the Ga-polar configuration. *(Note: The vertical arrows in (**B**) and (**C**) denote the absolute directions of the $P_{sp}$, which are corroboratively assigned based on subsequent PFM characterizations and theoretical calculations, further validating the geometric inversion observed herein.)*

To fundamentally elucidate the structural inversion and ascertain its crystallographic origin at the individual atomic-site level, we explicitly mapped the polyhedral configurations within the ε-$Ga_2O_3$ unit cells. As illustrated in the projected atomic model (Fig. 3a) and the corresponding annotated STEM image (Fig. 3b), the Ga-polar lattice is constructed by a network of three distinct gallium coordination environments: pentahedral (marked by blue circles), octahedral (red circles), and tetrahedral (green upward triangles) sites. The spatial arrangement of these multi-coordinated sites provides a direct visual signature of the internal crystal polarity. Crucially, while the pentahedral and octahedral columns establish the foundational backbone of the orthorhombic phase, it is the specific orientation of the tetrahedral polyhedra that dictates the non-centrosymmetric symmetry. In this configuration, the vertices of the tetrahedral polyhedra are strictly and uniformly oriented upwards, defining the lattice as the Ga-polar structure. In contrast, the atomic-site mapping of the O-polar film (Fig. 3c and d) reveals a comprehensive geometric inversion of this specific polyhedral network. High-resolution STEM confirms that the foundational atomic columns corresponding to the pentahedral and octahedral sites remain structurally intact, preserving the macroscopic orthorhombic framework. However, the localized tetrahedral sites (now marked by green downward triangles) exhibit a complete 180° structural flip. This rigorous intra-unit-cell polyhedral inversion perfectly corroborates the previously measured 0.12 nm atomic displacement along the c-axis (as shown in Fig. 2h). Consequently, this atomic-level geometric reversal provides the definitive structural origin for the macroscopic inversion of the $P_{sp}$ vectors between the two engineered heteroepitaxial architectures.

## 2.3 Polarity mapping and piezoelectric response

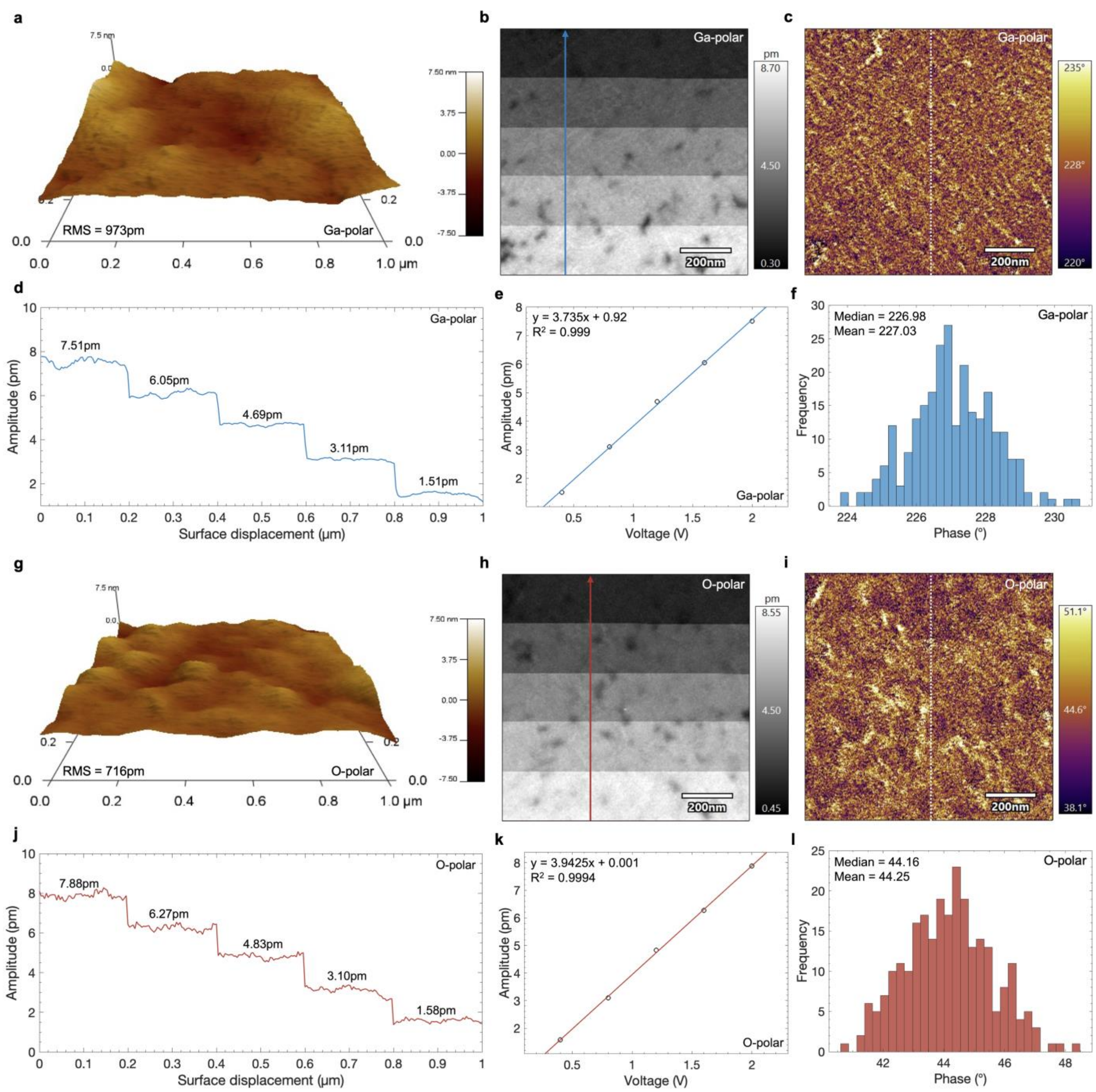

**Fig. 4 | Surface morphology and macroscopic piezoelectric response of the polarity-inverted ε-$Ga_2O_3$ films.** (**a** to **f**) Characterization of the Ga-polar film via DART-PFM. (**a**) 3D surface topography with an RMS roughness of 973 pm. (**b**) PFM amplitude map acquired under stepped AC drive voltages. (**c**) Corresponding PFM phase map. (**d**)

Amplitude profile extracted along the blue line in (**b**). (**e**) Linear fit of the average amplitude versus applied AC voltage, yielding an effective piezoelectric coefficient ($d_{33}$) of 3.735 pm/V. (**f**) Histogram of the phase distribution extracted from (**c**), centered at a mean phase angle of ~227°. (**g** to **l**) Corresponding characterization of the O-polar film. (**g**) 3D surface topography with an RMS roughness of 716 pm. (**h**) PFM amplitude map under stepped AC voltages and (**i**) the corresponding phase map. (**j**) Amplitude profile extracted along the red line in (**h**). (**k**) Linear fit yielding a comparable effective $d_{33}$ of 3.943 pm/V. (**l**) Phase distribution histogram centered at ~44°. *(Note: All PFM measurements were conducted with a strictly fixed instrumental phase offset. The resultant phase difference of ~183° between the two samples confirms an anti-parallel* $P_{sp}$ *orientation. However, DART-PFM cannot determine the absolute upward or downward direction of these* $P_{sp}$ *vectors.)*

To rigorously quantify the macroscopic piezoelectric response and map the domain orientations of the polarity-inverted films, dual AC resonance tracking piezoelectric force microscopy (DART-PFM) was employed. Before analyzing the data, the fundamental limitations of the instrumentation must be strictly defined to eliminate measurement artifacts. Because standard optical beam deflection (OBD) detection systems merely read cantilever deflection, the intrinsic piezo-signals are dynamically amplified by the cantilever resonance. Consequently, all amplitude data presented herein were rigorously extracted via simple harmonic oscillator (SHO) fitting. Furthermore, inherent instrumental phase offsets render isolated absolute phase values physically meaningless. To establish a reliable comparative baseline, a strictly identical instrumental phase offset was enforced during the measurements of both the Ga-polar and O-polar samples, ensuring that any resultant relative phase shift is exclusively attributed to the intrinsic crystal polarity.

Fig. 4 presents the macro-scale topographic and piezoelectric characteristics. The 3D surface topographies confirm highly smooth morphologies for both the Ga-polar (Fig. 4a, RMS roughness = 973 pm) and O-polar (Fig. 4g, RMS roughness = 716 pm) architectures, ruling out severe roughness-induced phase crosstalk. Under stepped AC drive voltages, the PFM amplitude maps (Fig. 4b and 4h) and their corresponding extracted profiles (Fig. 4d and 4j) exhibit distinct, discrete step-responses. The amplitude exhibits a highly linear dependence on the excitation voltage. Linear regression yields an effective piezoelectric coefficient ($d_{33}$) of 3.735 pm/V for the Ga-polar film (Fig. 4e) and a comparable value of 3.943 pm/V for the O-polar film (Fig. 4k). These robust and highly symmetric $d_{33}$ parameters quantitatively confirm the excellent macroscopic piezoelectric activity intrinsic to the ε-$Ga_2O_3$ orthorhombic phase. Crucially, the corresponding PFM phase maps (Fig. 4c and 4i) unveil the fundamental inversion of the $P_{sp}$. Statistical phase distribution histograms extracted from these maps demonstrate a highly concentrated mean phase angle of ~227° for the Ga-polar lattice (Fig. 4f) and ~44° for the O-polar lattice (Fig. 4l). The resultant phase difference of approximately 183° provides evidence of a completely anti-parallel orientation of the $P_{sp}$ vectors between the two engineered systems. However, constrained by the OBD-PFM operating principles, this ~180° relative phase inversion merely confirms an anti-parallel relationship; it is mathematically incapable of decoupling the absolute upward or downward physical direction of these $P_{sp}$ vectors.

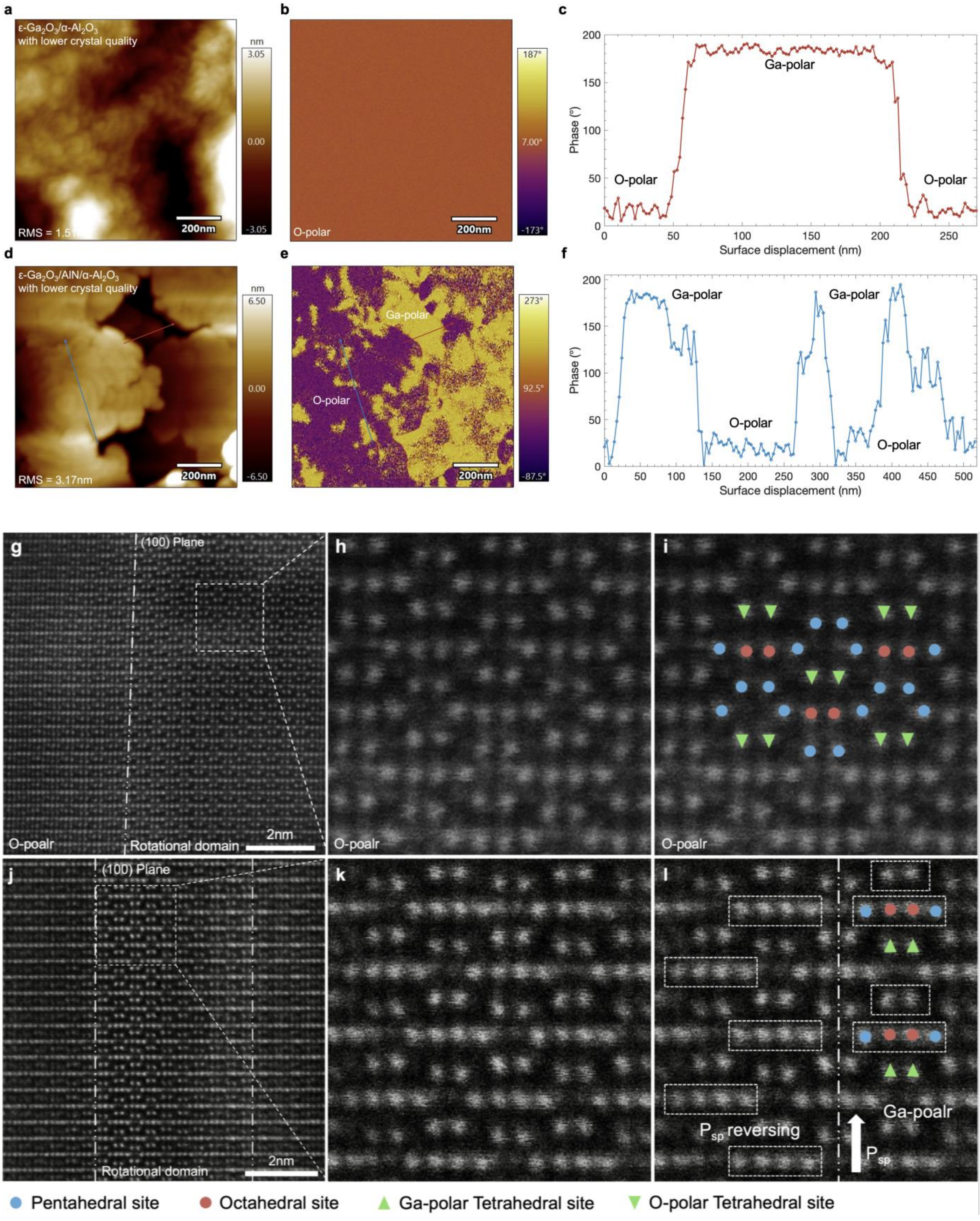

**Fig. 5 | Absolute polarity determination and atomic-scale domain architectures resolved by Vero interferometric PFM and STEM.** (**a** to **f**) Nanoscale mapping of the absolute $P_{sp}$ direction using interferometric PFM enabled by QPDI. (**a**) Surface topography (RMS roughness = 1.51 nm) and (**b**) corresponding phase map of the $\varepsilon$-$Ga_2O_3$/$\alpha$-$Al_2O_3$ sample with lower crystal quality, exhibiting a uniformly stable O-polar matrix. (**d**) Surface topography (RMS roughness = 3.17 nm) and (**e**) corresponding phase map of the $\varepsilon$-$Ga_2O_3$/AlN/$\alpha$-$Al_2O_3$ heterostructure with lower crystal quality, revealing the spatial coexistence of Ga-polar (yellow) and O-polar (purple)

sections. (**c** and **f**) Phase profiles extracted along the red and blue lines indicated in (**d**) and (**e**), respectively. By directly recording true out-of-plane surface displacement, the interferometric signals assign lattice expansion (phase ≈ 0°) to the O-polar regions and lattice contraction (phase ≈ 180°) to the Ga-polar domains. (**g** to **i**) Atomic-scale structural characterization of the lower crystal quality ε-$Ga_2O_3$/α-$Al_2O_3$ sample. (**g**) STEM image across a rotational domain. (**h**) Unannotated magnified view for pristine visual assessment. (**i**) Annotated identical view mapping the pentahedral (blue circles), octahedral (red circles), and tetrahedral (green downward triangles) sites, confirming strict O-polarity. (**j** to **l**) Atomic-scale validation of the mixed-polarity boundary in the lower crystal quality ε-$Ga_2O_3$/AlN/α-$Al_2O_3$ sample. (**j**) Cross-sectional STEM image capturing a distinct $P_{sp}$ inversion boundary. (**k**) Unannotated magnified view of the interfacial region. (**l**) Annotated identical view delineating the structural transition, with overlaid markers explicitly identifying the Ga-polar tetrahedral sites (green upward triangles) adjacent to the $P_{sp}$ reversing boundary.

To unambiguously decouple the absolute polarization vectors from instrumental phase artifacts, we deployed Vero PFM enabled by QPDI. Standard OBD systems are fundamentally limited to measuring cantilever buckling, whereas QPDI utilizes a dual-beam interferometric configuration targeting both the probe base (reference) and the tip to directly quantify true out-of-plane surface displacement. This configuration eliminates arbitrary phase offsets. Within this physical framework, a measured phase near 0° denotes lattice expansion (occurring when the downward tip electric field aligns parallel to a downward $P_{sp}$), while a phase near 180° indicates lattice contraction (downward tip electric field aligns anti-parallel to an upward $P_{sp}$).

Applying this criterion, we evaluated the lower-crystal-quality samples. The macroscopic structural conditions of these specific films were first quantitatively assessed via XRD (fig. S3). While the 2θ scans (fig. S3a and 3c) confirm the rigorous preservation of phase purity without any parasitic secondary phases, the corresponding (004) rocking curves (fig. S3b and 3d) indicate a reduction in crystalline quality. Specifically, the FWHM of the ε-$Ga_2O_3$ film on the AlN template broadened to 0.36° (compared to 0.16° for the high-quality counterpart), and the FWHM of the film directly on α-$Al_2O_3$ broadened to 0.16° (compared to 0.09°). Despite this macroscopic crystalline degradation, for the lower-quality ε-$Ga_2O_3$ film grown directly on α-$Al_2O_3$ (Fig. 5a and 5b), the QPDI mapping exhibits an uniform interferometric phase concentrated at ~7°. This ~0° response assigns a downward absolute $P_{sp}$ vector, confirming a stable O-polar matrix. Atomic-scale STEM mapping (Fig. 5g to 5i) corroborates this macroscopic continuity. Despite the reduced overall crystal quality, the localized unit cells preserve the characteristic O-polar architecture. Specifically, the intra-cell tetrahedral sites (highlighted by green downward triangles in Fig. 5i) maintain their 180°-inverted geometry relative to the Ga-polar state, proving that the substrate-determined downward polarity is highly resilient to macroscopic lattice degradation.

Conversely, the ε-$Ga_2O_3$ film deposited on the AlN template with lower-crystal-quality (Fig. 5c to 5f) reveals a spatial coexistence of Ga-polar (phase ~180°, yellow domains) and O-polar (phase ~0°, purple domains) regions. Phase profiles extracted across these heterogeneous regions (Fig. 5c and 5f) explicitly capture the sharp, binary nature of this polarity inversion. Crucially, the spatial distribution of these inverted domains exhibits no morphological correlation with the random epitaxial island boundaries observed in the topography channel (Fig. 5d). This geometric independence eliminates the possibility of topography-induced phase crosstalk, isolating the signal as a genuine intrinsic structural inversion. This macroscopic phase coexistence is validated at the atomic scale (Fig. 5j to 5l). High-resolution STEM viewing along the (100) projection captures a distinct $P_{sp}$ inversion boundary. In the right half of the mapped region (Fig. 5l, Ga-polar region), the characteristic four-atom sequence exhibits the signature central elevation. However, the pentahedral sites directly above them fundamentally deviate from parallel alignment (indicated by the small dashed boxes in Fig. 5l). Moving across the boundary to the left half, this four-atom sequence transitions into a rigidly horizontal configuration (large dashed boxes), contrasting with the pristine Ga-polar lattice previously established (Fig. 3b). This severe intra-lattice structural distortion provides direct, atomistic evidence for the coexisting Ga-polar and O-polar architectures mapped by QPDI PFM.

To further investigate the nature of these polar sections, high-voltage Vero switching spectroscopy PFM (SS-PFM) was executed across all synthesized samples. Up to extreme tip biases of 120 V, no ferroelectric switching was induced in any configuration (including the high-quality epitaxial films on both AlN and $\alpha$-$Al_2O_3$, the lower-crystal-quality film on $\alpha$-$Al_2O_3$, and specifically both the Ga-polar and O-polar segregated micro-domains within the lower-quality AlN-templated heterostructure). This confirms that the observed domains possess rigidly pinned structural polarities rather than switchable ferroelectric states. We attribute this deterministic polarity to substrate-induced epitaxial registry, a phenomenon widely recognized in III-V semiconductor heteroepitaxy (e.g., Metal-polar vs. N-polar control). In pristine growths, the established upward $P_{sp}$ of the Al-polar AlN template dynamically induces a coherent upward $P_{sp}$ (Ga-polar) in the epilayer. However, under lower-quality growth conditions, the localized accumulation of epitaxial strain and extended defects severely screens the interfacial polarization field of the AlN template. This localized electrostatic screening disrupts the coherent transfer of polarity, triggering a spontaneous structural relaxation into the thermodynamically competitive O-polar state, ultimately yielding the nanoscale mixed-polarity architectures.

## 2.4 Theoretical origin of spontaneous polarization and piezoelectricity

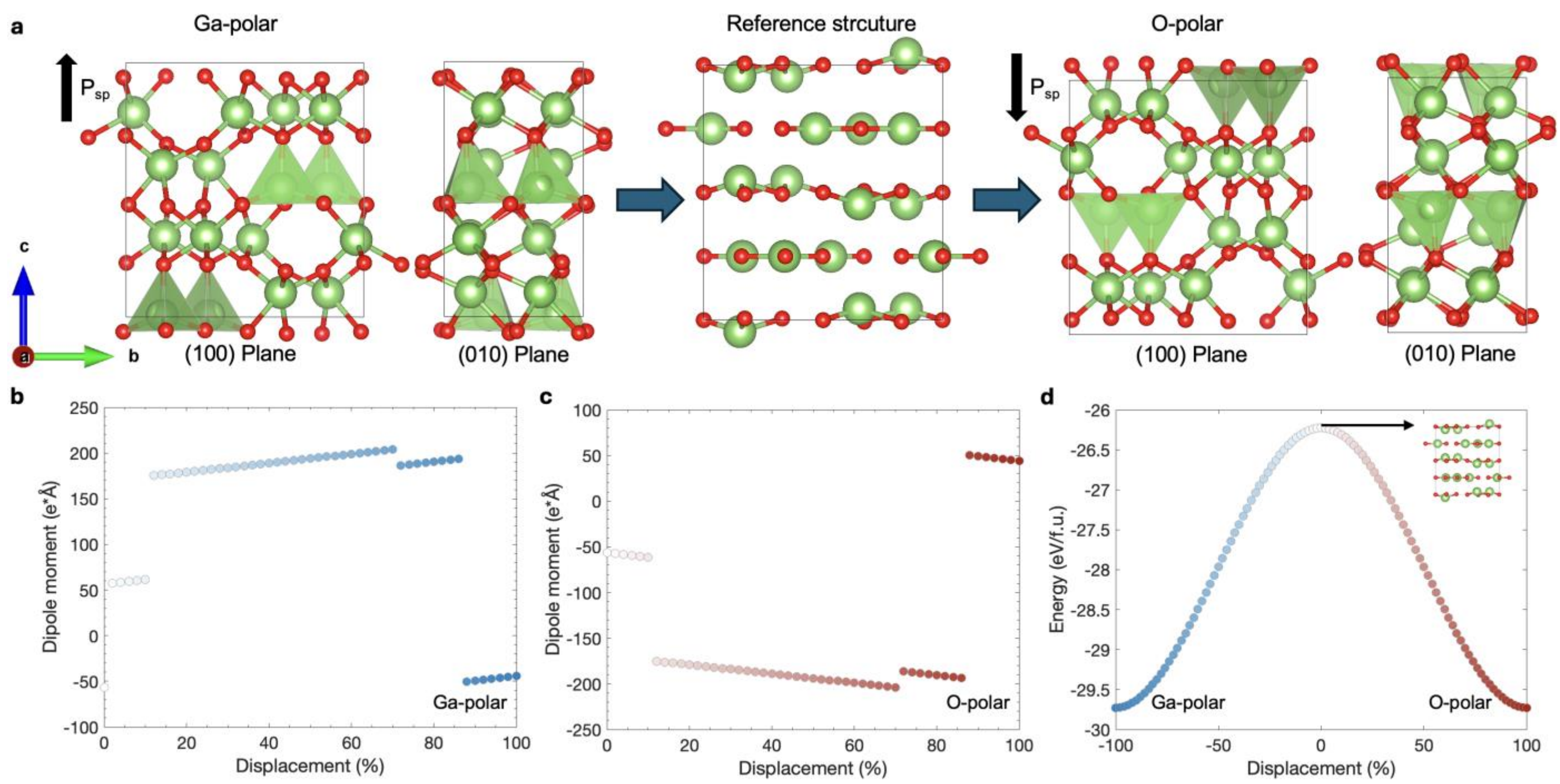

**Fig. 6 | First-principles evaluation of absolute spontaneous polarization and transition energetics in ε-$Ga_2O_3$.** (**a**) Atomistic pathways mapping the polar ε-$Ga_2O_3$ lattices (Ga-polar and O-polar) to a theoretically constructed centrosymmetric reference structure. The distinct tetrahedral orientations are visualized along the (100) and (010) planes, with vertical arrows denoting the mathematically derived absolute $P_{sp}$ vectors. (**b** and **c**) Evolution of the dipole moment evaluated from centrosymmetric reference structure to (**b**) Ga-polar and (**c**) O-polar configurations across different transition states. (**d**) Calculated total energy profile (in eV/f.u.) as a function of the structural displacement transitioning from the Ga-polar (-100%) to the O-polar (100%) state via the centrosymmetric reference structure at 0% displacement.

In accordance with the modern theory of polarization, evaluating the absolute $P_{sp}$ requires referencing a centrosymmetric phase [27]. Utilizing the Bilbao Crystallographic Server, a non-polar Pnma structure was strictly defined as the centrosymmetric reference (Fig. 6a, center) [28]. The two opposing polar $Pna2_1$ states (Ga-polar and O-polar) are depicted on either side, characterized by their structurally inverted tetrahedral coordination polyhedra. Critically, for all computational frameworks herein, the positive c-axis of the coordinate system is rigidly anchored parallel to the physical $P_{sp}$ direction of the Ga-polar phase (and thus anti-parallel to the O-polar $P_{sp}$).

Direct comparison of dipole moments between symmetry-distinct configurations is physically invalid due to the polarization quantum effect [27]. To overcome this ambiguity, 50 intermediate transition states were generated for each transformation pathway (from the reference to the Ga-polar and O-polar states, respectively). Fig. 6b and 6c map the evolution of the total dipole moment, comprising both the ionic point-charge contribution and electronic contribution derived via the Berry phase formulation. The data points transition from light to dark, visually encoding the structural deviation from the centrosymmetric origin. This interpolation smoothly resolves the multiple parallel branches in the dipole moment landscape, isolating the distinct polarization quantum states. The $P_{sp}$ magnitude is subsequently derived from the quantized dipole moment difference between the polar $Pna2_1$ states and the reference (which inherently dictates a zero dipole moment by symmetry), normalized by the unit cell volume. For the Ga-polar phase (Fig. 6b), the calculated $P_{sp}$ is +23.12 μC/cm$^2$. The positive slope explicitly confirms that the polarization vector aligns with the positive c-axis (tetrahedral vertices pointing upward). This theoretical alignment corroborates our prior Vero PFM measurements: a phase near 180° corresponds to lattice contraction under a downward tip electric field, physically requiring an upward-pointing absolute $P_{sp}$. Conversely, the evaluation for the O-polar phase (Fig. 6c) yields an identical magnitude but an inverted trajectory, calculating a $P_{sp}$ of -23.12 μC/cm2. This negative sign mathematically locks the O-polar $P_{sp}$ anti-parallel to the c-axis (tetrahedral vertices pointing downward), matching the Vero PFM response where a ~0° phase indicates lattice expansion under a downward electric field. Furthermore, Fig. 6d delineates the calculated total energy profile transitioning from the Ga-polar state (-100%) through the centrosymmetric reference (0%) to the O-polar state (100%). The resulting continuous energy landscape verifies the structural validity of the generated intermediate states.

To translate these polarization dynamics into electromechanical responses, the piezoelectric stress tensor (*e*) and the elastic stiffness tensor (*C*) were computed for both polarities (values detailed in table S1-S4). The calculations reveal that the elastic constants remain identical between the two phases. This is a strict physical mandate, as elasticity is governed by a fourth-rank tensor that is invariant under spatial inversion. In contrast, the piezoelectric stress tensor , being a third-rank tensor, dictates an odd-parity response and must rigorously flip its sign upon inversion of the polar axis. The $d_{33}$ was subsequently extracted via the fundamental relationship $d=eC^{-1}$. For the Ga-polar state, the calculated $d_{33}$ is +5.034 pm/V, which closely mirrors the experimentally measured effective $d_{33}$ of 3.735 pm/V. This positive $d_{33}$ physically dictates that the lattice will expand under an electric field applied along the positive c-axis (upward), and geometrically contract under a negative c-axis field (downward). This closes the logical loop with the experimental Vero-PFM setup, where a downward drive field induced lattice contraction (180° phase). Correspondingly, the O-polar state yields a $d_{33}$ of -4.976 pm/V. This negative coefficient physically guarantees lattice contraction under an upward field and expansion under a downward field, impeccably mirroring the ~0° phase experimental mapping.

## 3. Conclusion

In summary, our atomic-to-macroscopic correlation framework resolves the fundamental ambiguity surrounding the absolute $P_{sp}$ vectors in polar ε-$Ga_2O_3$. By establishing the asymmetric four-atom geometric sequence as a structural criterion, we demonstrate that the macroscopic $P_{sp}$ direction can be governed through precise interface registry with Al-polar AlN and sapphire substrates. Furthermore, we reveal that deteriorated crystalline quality disrupts this substrate registry, leading to thermodynamically pinned, mixed-polarity domains. Successfully preventing such mixed-polarity relaxation allows for the synthesis of uniformly Ga-polar and O-polar matrices, unlocking the intentional design of macroscopic polarization discontinuities at oxide heterointerfaces. Analogous to mature III-Nitride architectures, this capability provides the essential foundational platform for inducing high-density, dopant-free 2DEG and tailoring internal electric fields. Consequently, this substrate-engineered blueprint transitions ε-$Ga_2O_3$ into a highly viable, tunable polar semiconductor for next-generation power and radio-frequency electronics.

# 4. Methods

## 4.1 Materials epitaxy

Prior to deposition, substrates were subjected to sequential ultrasonic cleaning in acetone, de-ionized (DI) water, and isopropanol (IPA). The epitaxial growth was conducted using a low-pressure mist chemical vapor deposition (LP-Mist-CVD) system [29, 30]. The system utilizes a hot-wall tubular furnace with a 150 mm diameter and 1500 mm length, divided into three distinct temperature zones: zone 1 at 550 °C, zone 2 at 400 °C, and zone 3 at 250 °C. Specifically, Ga-polar and O-polar ε-$Ga_2O_3$ films were grown on commercial Al-polar AlN/sapphire (2μm thickness of AlN) and c-plane sapphire substrates, respectively. The gallium precursor was a 0.04 M aqueous solution of $Ga(acac)_3$ (Macklin, 99.99%), which was stabilized by acidification with 3.7 vol% HCl (Guangzhou Chemical Reagent Factory, 36.0–38.0%). The precursor mist was generated using an ultrasonic transducer operating at 2.4 MHz. Nitrogen ($N_2$) was utilized as the carrier gas with flow rates strictly configured at 3000 sccm for the mist line and 1000 sccm for the mist purge line. Pure oxygen ($O_2$) was concurrently introduced at a flow rate of 1000 sccm. During the process, the furnace operating pressure was dynamically regulated to 850 mbar using a vacuum pump to guarantee consistent mist delivery and uniform deposition. The substrates were positioned in zone 1 (550 °C) for 1 hour to initiate epitaxial growth within the region of maximum precursor concentration. This controlled deposition yielded a steady growth rate of ~200 nm/h, resulting in a uniform final film thickness of approximately 200 nm across all characterized samples.Upon completion of the deposition period, the substrate holder was slowly translated from zone 1 to zone 3 to execute a rapid yet controlled cooling sequence, effectively quenching the chemical reaction.

## 4.2 Characterization

To ascertain the phase composition and crystallographic properties of the samples, XRD was performed utilizing a Malvern PANalytical Empyrean 3.0 system equipped with a PIXcel3D detector. Cross-sectional samples were extracted via a $Ga^+$ FIB in a Thermo Fisher Helios 5 UX dual-beam system. Subsequent atomic-scale microstructural investigations were conducted on a spherical aberration-corrected transmission electron microscope (JEM-ARM200F NEOARM) operating in HAADF-STEM mode, supplemented by EDS for precise chemical mapping.

Surface morphology and piezoelectric behaviors were evaluated using an Oxford Instruments Cypher VRS1250 AFM deployed in dual AC resonance tracking switching spectroscopy piezoelectric force microscopy (DART-SS-PFM) mode. Furthermore, to accurately resolve the spontaneous polarization directions in samples exhibiting inferior crystalline quality, high-resolution mapping was executed on an Oxford Instruments Vero interferometric PFM utilizing QPDI technology. This interferometric approach directly measures true out-of-plane surface displacement, unambiguously distinguishing between lattice expansion (phase near 0°) and contraction (phase near 180°) to eliminate artifact-induced phase ambiguities. Both the topography and local piezoresponse were mapped utilizing commercially available conductive Pt-coated silicon probes (HQ:NSC18/Pt, μmasch) with a nominal resonance frequency of 75 kHz and a force constant of 2.8 N/m.

## 4.3 DFT calculation

First-principles density functional theory (DFT) calculations were executed using the Vienna Ab initio Simulation Package (VASP) [31]. Electron-ion interactions were described by projector-augmented wave (PAW) pseudopotentials, and the exchange-correlation functional was parameterized via the Perdew-Burke-Ernzerhof (PBE) generalized gradient approximation (GGA) [32-34]. To guarantee rigorous structural convergence, the plane-wave kinetic energy cutoff was strictly fixed at 650 eV. $P_{sp}$ was evaluated via the Berry phase approach within modern polarization theory [27]. The prerequisite symmetry analysis, mapping ε-$Ga_2O_3$ to its reference centrosymmetric phase, was conducted using the Bilbao Crystallographic Server [28].

## Acknowledgements

The work was supported by C. K. Tan start-up fund from the Hong Kong University of Science and Technology (Guangzhou); Guangzhou Municipal Science and Technology Project (No. 2023A03J0003, No. 2023A03J0013, No. 2023A04J0310 and No. 2023A03J0152); Department of Education of Guangdong Province (No. 2024ZDZX1005); State Administration of Foreign Experts Affairs (No. Y20240005); Matching Funding for Selected Talent of National Programs (CZ118SC24007); National Major Talent Project (CZ118SC25005); Excellent Young Scientists Fund (overseas) (RK118QN24006). This work was supported by the Materials Characterization and Preparation Facility (MCPF) and Green Materials Laboratory at the Hong Kong University of Science and Technology (Guangzhou).

## References

1. Pearton, S. J.; Yang, J.; Cary, P. H., IV; Ren, F.; Kim, J.; Tadjer, M. J.; Mastro, M. A., A review of Ga2O3 materials, processing, and devices. *Applied Physics Reviews* **2018,** *5* (1).
2. Higashiwaki, M.; Sasaki, K.; Murakami, H.; Kumagai, Y.; Koukitu, A.; Kuramata, A.; Masui, T.; Yamakoshi, S., Recent progress in Ga2O3 power devices. *Semiconductor Science and Technology* **2016,** *31* (3), 034001.
3. Higashiwaki, M.; Sasaki, K.; Kuramata, A.; Masui, T.; Yamakoshi, S., Gallium oxide (Ga2O3) metal-semiconductor field-effect transistors on single-crystal β-Ga2O3 (010) substrates. *Applied Physics Letters* **2012,** *100* (1).
4. Higashiwaki, M.; Sasaki, K.; Kamimura, T.; Hoi Wong, M.; Krishnamurthy, D.; Kuramata, A.; Masui, T.; Yamakoshi, S., Depletion-mode Ga2O3 metal-oxide-semiconductor field-effect transistors on β-Ga2O3 (010) substrates and temperature dependence of their device characteristics. *Applied Physics Letters* **2013,** *103* (12).
5. He, H.; Orlando, R.; Blanco, M. A.; Pandey, R.; Amzallag, E.; Baraille, I.; Rérat, M., First-principles study of the structural, electronic, and optical properties of ${\mathrm{Ga}}_{2}{\mathrm{O}}_{3}$ in its monoclinic and hexagonal phases. *Physical Review B* **2006,** *74* (19), 195123.
6. Maccioni, M. B.; Fiorentini, V., Phase diagram and polarization of stable phases of (Ga1-xInx)2O3. *Appl. Phys. Express* **2016,** *9*, 041102.
7. Kim, J.; Tahara, D.; Miura, Y.; Kim, B. G., First-principle calculations of electronic structures and polar properties of (κ,ε)-Ga2O3. *Applied Physics Express* **2018,** *11* (6), 061101.
8. Ranga, P.; Cho, S. B.; Mishra, R.; Krishnamoorthy, S., Highly tunable, polarization-engineered two-dimensional electron gas in ε-AlGaO3/ε-Ga2O3 heterostructures. *Applied Physics Express* **2020,** *13* (6), 061009.
9. Cho, S. B.; Mishra, R., Epitaxial engineering of polar ε-Ga2O3 for tunable two-dimensional electron gas at the heterointerface. *Applied Physics Letters* **2018,** *112* (16), 2101.
10. Wang, Y.; Cao, J.; Song, H.; Zhang, C.; Xie, Z.; Wong, Y. H.; Tan, C. K., Polarization engineering of two-dimensional electron gas at ε-(AlxGa1–x)2O3/ε-Ga2O3 heterostructure. *Applied Physics Letters* **2023,** *123* (14).
11. Khan, M. A.; Kuznia, J. N.; Van Hove, J. M.; Pan, N.; Carter, J., Observation of a two-dimensional electron gas in low pressure metalorganic chemical vapor deposited GaN-AlxGa1−xN heterojunctions. *Applied Physics Letters* **1992,** *60* (24), 3027-3029.
12. Toyama, S.; Seki, T.; Kanitani, Y.; Kudo, Y.; Tomiya, S.; Ikuhara, Y.; Shibata, N., Real-space observation of a two-dimensional electron gas at semiconductor heterointerfaces. *Nature Nanotechnology* **2023,** *18* (5), 521-528.
13. Mishra, U. K.; Parikh, P.; Yi-Feng, W., AlGaN/GaN HEMTs-an overview of device operation and applications. *Proceedings of the IEEE* **2002,** *90* (6), 1022-1031.

14. Simon, J.; Protasenko, V.; Lian, C.; Xing, H.; Jena, D., Polarization-Induced Hole Doping in Wide–Band-Gap Uniaxial Semiconductor Heterostructures. *Science* **2010,** *327* (5961), 60-64.
15. Chaudhuri, R.; Bader, S. J.; Chen, Z.; Muller, D. A.; Xing, H. G.; Jena, D., A polarization-induced 2D hole gas in undoped gallium nitride quantum wells. *Science* **2019,** *365* (6460), 1454-1457.
16. Ambacher, O.; Smart, J.; Shealy, J. R.; Weimann, N. G.; Chu, K.; Murphy, M.; Schaff, W. J.; Eastman, L. F.; Dimitrov, R.; Wittmer, L.; Stutzmann, M.; Rieger, W.; Hilsenbeck, J., Two-dimensional electron gases induced by spontaneous and piezoelectric polarization charges in N- and Ga-face AlGaN/GaN heterostructures. *Journal of Applied Physics* **1999,** *85* (6), 3222-3233.
17. Rajan, S.; Chini, A.; Wong, M. H.; Speck, J. S.; Mishra, U. K., N-polar GaN∕AlGaN∕GaN high electron mobility transistors. *Journal of Applied Physics* **2007,** *102* (4).
18. Wong, M. H.; Keller, S.; Dasgupta, N. S.; Denninghoff, D. J.; Kolluri, S.; Brown, D. F.; Lu, J.; Fichtenbaum, N. A.; Ahmadi, E.; Singisetti, U.; Chini, A.; Rajan, S.; DenBaars, S. P.; Speck, J. S.; Mishra, U. K., N-polar GaN epitaxy and high electron mobility transistors. *Semiconductor Science and Technology* **2013,** *28* (7), 074009.
19. van Deurzen, L.; Kim, E.; Pieczulewski, N.; Zhang, Z.; Feduniewicz-Zmuda, A.; Chlipala, M.; Siekacz, M.; Muller, D.; Xing, H. G.; Jena, D.; Turski, H., Using both faces of polar semiconductor wafers for functional devices. *Nature* **2024,** *634* (8033), 334-340.
20. Dreyer, C. E.; Janotti, A.; Van de Walle, C. G.; Vanderbilt, D., Correct Implementation of Polarization Constants in Wurtzite Materials and Impact on III-Nitrides. *Physical Review X* **2016,** *6* (2), 021038.
21. Fichtner, S.; Yassine, M.; Van de Walle, C. G.; Ambacher, O., Clarification of the spontaneous polarization direction in crystals with wurtzite structure. *Applied Physics Letters* **2024,** *125* (4).
22. Wang, D.; Wang, D.; Yang, S.; Mi, Z., Rethinking polarization in wurtzite semiconductors. *Applied Physics Letters* **2024,** *124* (26).
23. Kracht, M.; Karg, A.; Schörmann, J.; Weinhold, M.; Zink, D.; Michel, F.; Rohnke, M.; Schowalter, M.; Gerken, B.; Rosenauer, A.; Klar, P. J.; Janek, J.; Eickhoff, M., Tin-Assisted Synthesis of ε-Ga2O3 by Molecular Beam Epitaxy. *Phys. Rev. Appl.* **2017,** *8*, 054002.
24. Oshima, Y.; Kawara, K.; Oshima, T.; Shinohe, T., In-plane orientation control of (001) κ-Ga2O3 by epitaxial lateral overgrowth through a geometrical natural selection mechanism. *Japanese Journal of Applied Physics* **2020,** *59* (11), 115501.
25. Nishinaka, H.; Komai, H.; Tahara, D.; Arata, Y.; Yoshimoto, M., Microstructures and rotational domains in orthorhombic ε-Ga2O3 thin films. *Japanese Journal of Applied Physics* **2018,** *57* (11), 115601.
26. Ye, H.; Wang, P.; Wang, R.; Wang, J.; Xu, X.; Feng, R.; Wang, T.; Tong, W.-Y.; Liu, F.; Sheng, B.; Ma, W.; An, B.; Li, H.; Chen, Z.; Duan, C.-G.; Ge, W.; Shen, B.; Wang, X., Experimental determination of giant polarization in wurtzite III-nitride semiconductors. *Nature Communications* **2025,** *16* (1), 3863.
27. Spaldin, N. A., A beginner's guide to the modern theory of polarization. *Journal of Solid State Chemistry* **2012,** *195*, 2-10.
28. Capillas, C.; Tasci, E. S.; Flor, G. d. l.; Orobengoa, D.; Perez-Mato, J. M.; Aroyo, M. I. *A new computer tool at the Bilbao Crystallographic Server to detect and characterize pseudosymmetry*, 2011.
29. Wang, Y.; Xie, Z.; Guan, Y.; Zhang, C.; Zhang, Y.; Cao, J.; Ying, Z.; Hu, G.; Tan, C. K., Sn-assisted epitaxial growth of high-crystallinity κ/ε-Ga2O3 on sapphire (0001) by low-pressure Mist-CVD. *Applied Physics Letters* **2025,** *127* (10), 102109.
30. Wang, Y.; Xie, Z.; Guan, Y.; Cao, J.; Zhang, Y.; Hu, G.; Bai, Y.; Wong, Y. H.; Zeng, G.; Huang, Z.; Tan, C. K., Atomic-level revelation of spontaneous polarization orientation and piezoelectricity in ε-Ga2O3. *Applied Physics Letters* **2026,** *128* (12).
31. MedeA-VASP *MedeA-VASP, Material Designs Incwww.materialsdesign. com/medea/medea-vasp.*

32. Kresse, G.; Furthmüller, J., Efficient iterative schemes for ab initio total-energy calculations using a plane-wave basis set. *Physical Review B* **1996,** *54* (16), 11169-11186.
33. Blochl, P. E., Projector augmented-wave method. *Physcal Review B* **1994,** *50* (24), 17953.
34. Perdew, J. P.; Burke, K.; Ernzerhof, M., Generalized Gradient Approximation Made Simple. *Physical Review Letters* **1996,** *77* (18), 3865.

## Supporting information

# Epitaxial inversion of spontaneous polarization in ε-$Ga_2O_3$

**Yan Wang[1,2*], Zhigao Xie[1,2], Weihua Tang[1*], Chee Keong Tan[1,2,3,4,5*]**

*[1]College of Integrated Circuit Science and Engineering, Nanjing University of Posts and Telecommunications, Nanjing 210023, China*

*[2]Advanced Materials Thrust, Function Hub, The Hong Kong University of Science and Technology (Guangzhou), Guangzhou 511453, Guangdong, China*

*[3]Department of Electronic and Computer Engineering, The Hong Kong University of Science and Technology, Hong Kong, China*

*[4]Guangzhou Municipal Key Laboratory of Materials Informatics, The Hong Kong University of Science and Technology (Guangzhou), Guangzhou 511453, Guangdong, China*

*[5]Guangzhou Municipal Key Laboratory of Integrated Circuits Design, The Hong Kong University of Science and Technology (Guangzhou), Guangzhou 511453, Guangdong, China*

* Corresponding author.

Email addresses: ywang950@connect.hkust-gz.edu.cn (Y. Wang); whtang@njupt.edu.cn (W. Tang); 20230193@njupt.edu.cn & cheekeongtan@hkust-gz.edu.cn (C. K. Tan)

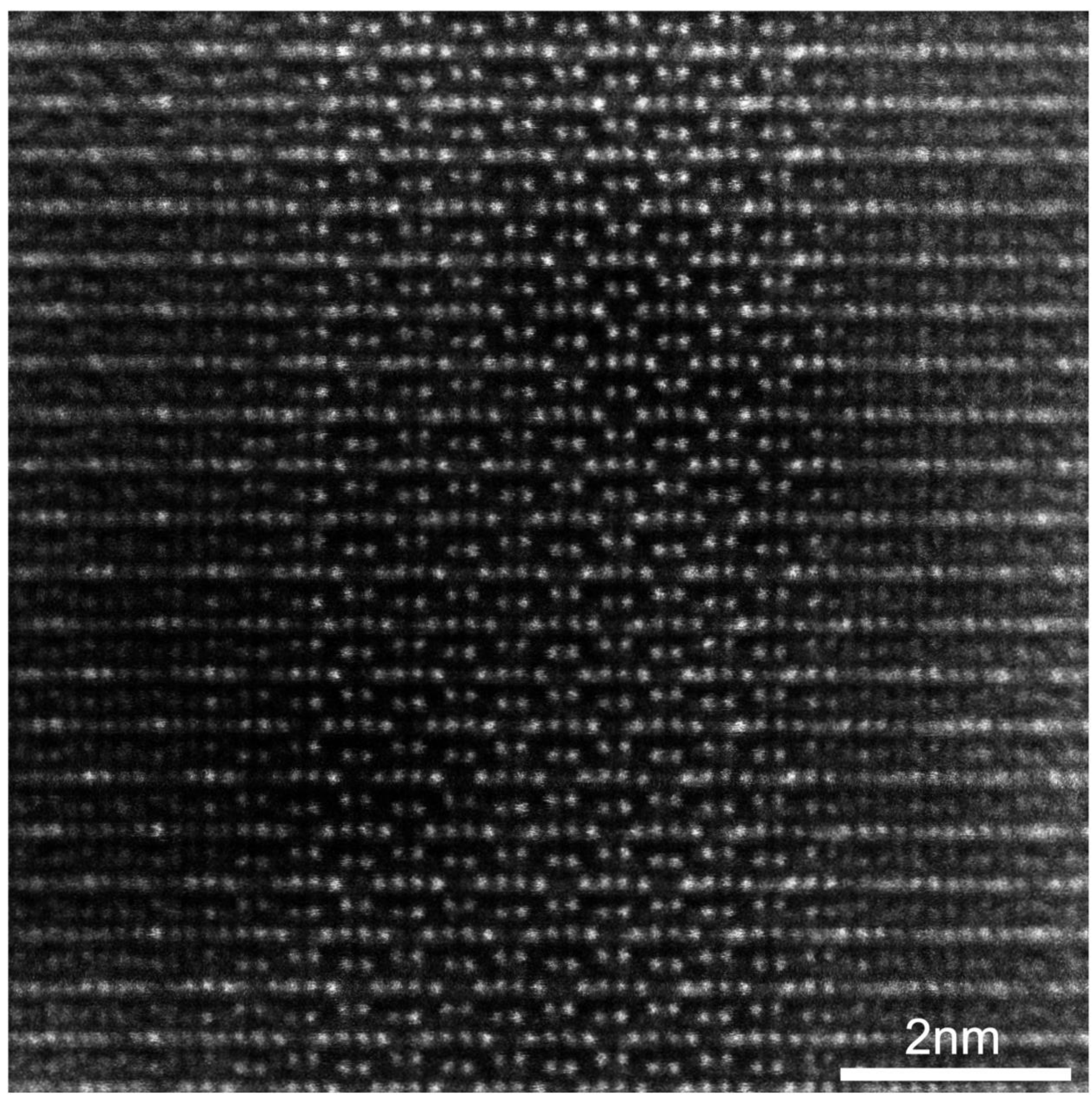

**Fig. S1 | High-resolution, large-scale STEM imaging of the Ga-polar ε-$Ga_2O_3$ lattice.** This figure provides a full-scale, high-resolution version of the HAADF-STEM image presented in Fig. 2b of the main text.

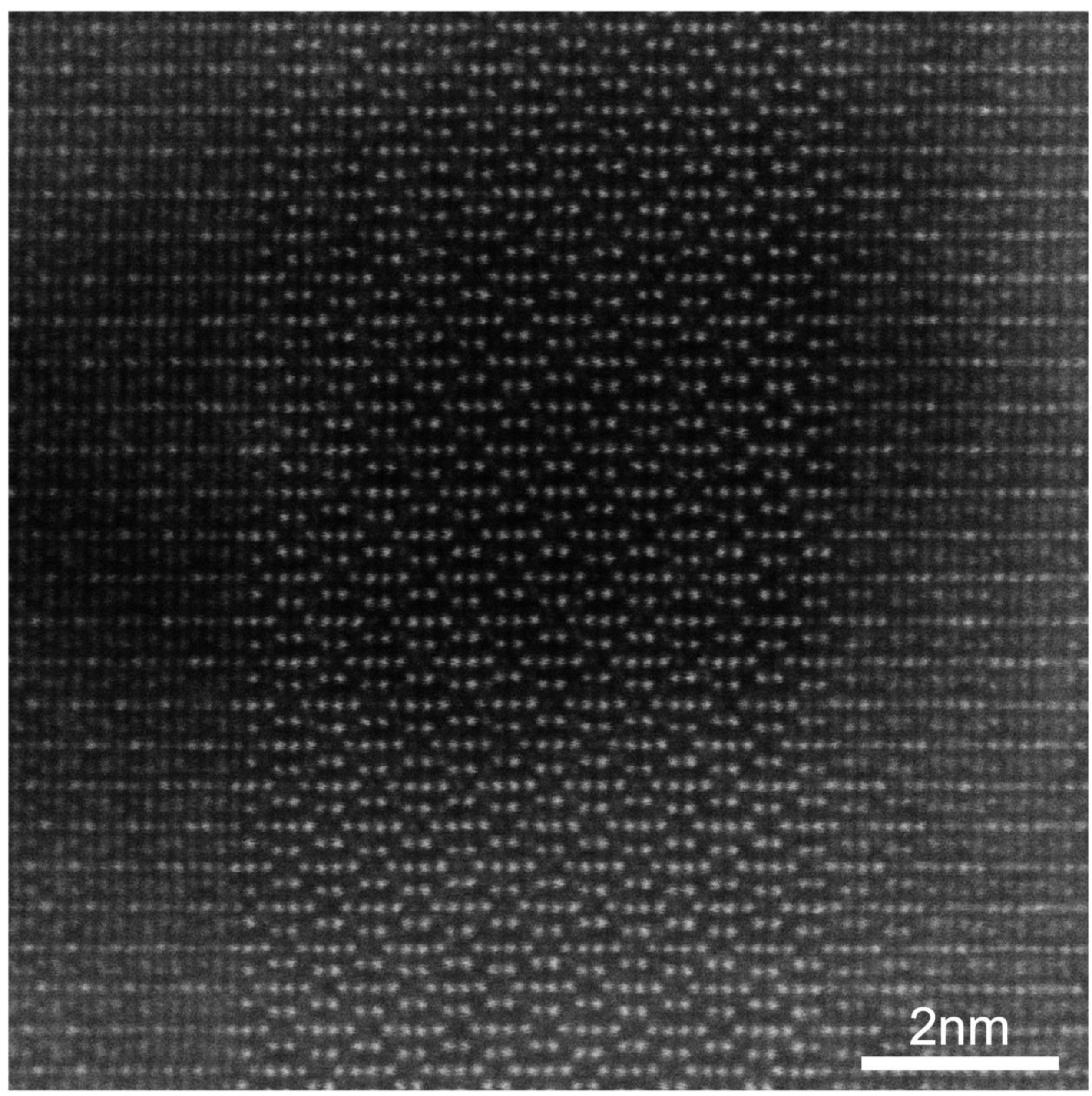

**Fig. S2 | High-resolution, large-scale STEM imaging of the O-polar ε-$Ga_2O_3$ lattice.** This figure provides a full-scale, high-resolution version of the HAADF-STEM image presented in Fig. 2g of the main text.

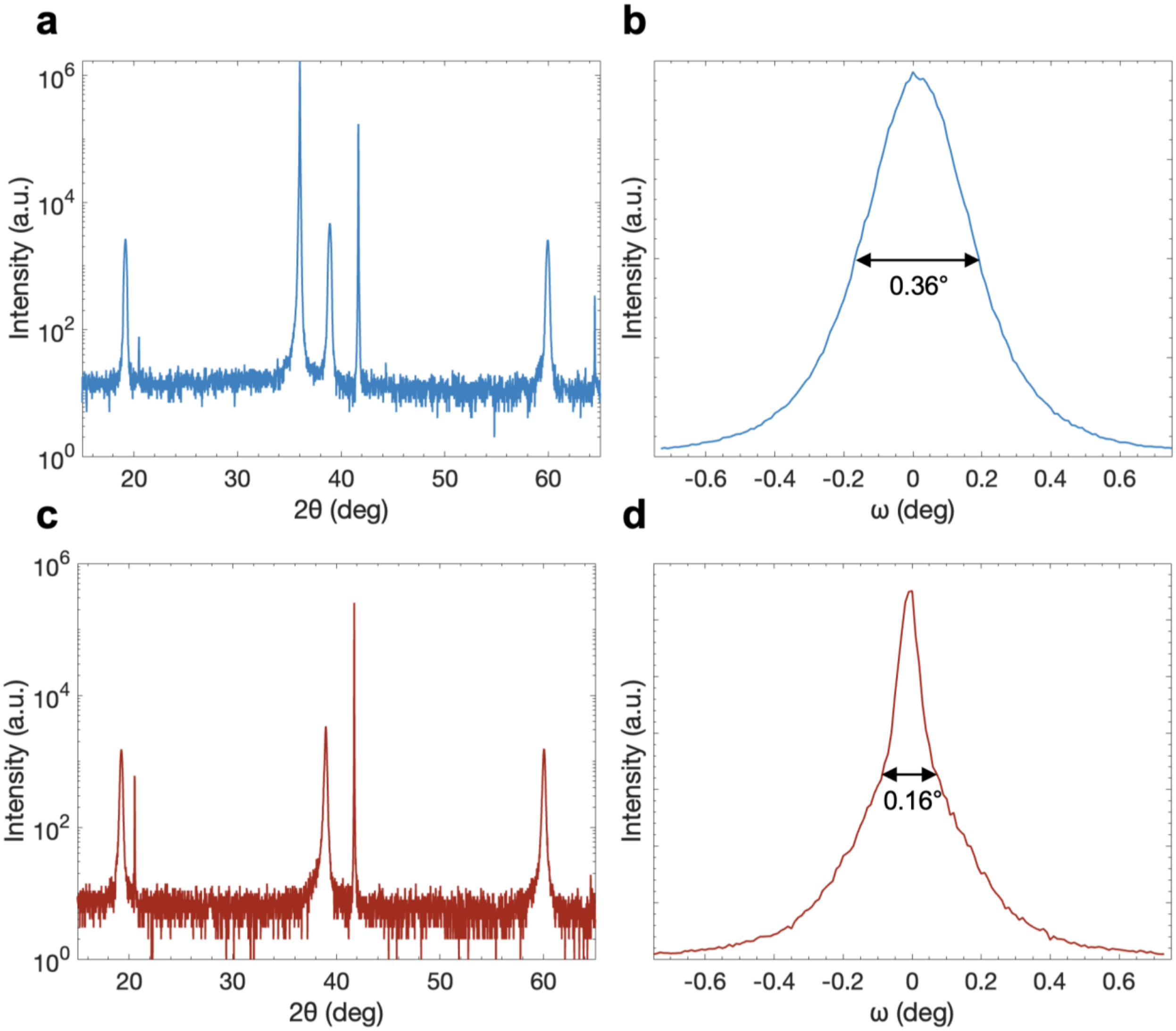

**Fig. S3 | Structural characterization of ε-$Ga_2O_3$ epitaxial films with reduced crystalline quality.** (**a** and **b**) XRD profiles of the lower-quality ε-$Ga_2O_3$/AlN/α-$Al_2O_3$ heterostructure, exhibiting (**a**) the 2θ-scan and (**b**) the ω-scan rocking curve of the ε-$Ga_2O_3$ (004) reflection (FWHM = 0.36°). (**c** and **d**) Corresponding profiles of the lower-quality ε-$Ga_2O_3$ film grown directly on the α-$Al_2O_3$ substrate, showing (**c**) the 2θ-scan and (**d**) the ω-scan rocking curve of the ε-$Ga_2O_3$ (004) reflection (FWHM = 0.16°).

**Table S1 | Piezoelectrc constant ($e$) for Ga-poalr ε-$Ga_2O_3$**

| | XX | YY | ZZ | XY | YZ | ZX |
|---|---|---|---|---|---|---|
| x | 0 | 0 | 0 | 0 | 0 | 1.1109 |
| y | 0 | 0 | 0 | 0 | 0.35347 | 0 |
| z | 0.18139 | -0.4961 | 0.84456 | 0 | 0 | 0 |

**Table S2 | Elastic constant ($C$) for Ga-poalr ε-$Ga_2O_3$**

| | XX | YY | ZZ | XY | YZ | ZX |
|---|---|---|---|---|---|---|
| XX | 285.6858 | 142.8693 | 114.5747 | -0.1701 | -0.1039 | 0.0749 |
| YY | 142.8693 | 252.0863 | 115.7865 | -0.1746 | -0.126 | 0.0965 |
| ZZ | 114.5747 | 115.7865 | 255.9722 | -0.1875 | -0.1272 | 0.0945 |
| XY | -0.1701 | -0.1746 | -0.1875 | 64.7837 | 0.0013 | -0.0014 |
| YZ | -0.1039 | -0.126 | -0.1272 | 0.0013 | 39.2734 | 0.0109 |
| ZX | 0.0749 | 0.0965 | 0.0945 | -0.0014 | 0.0109 | 85.1166 |

piezoelectric strain tensor ($d_{33}$) for Ga-poalr ε-$Ga_2O_3$:

$$d_{33} = \sum_{j=1}^{6} e_{3j}\,(C^{-1})_{j3}$$

$$d_{33} = e_{31}(C^{-1})_{13} + e_{32}(C^{-1})_{23} + e_{33}(C^{-1})_{33}$$

$$d_{33} = \ 5.034\ pm/V$$

**Table S3 | Piezoelectrc constant ($e$) for O-poalr ε-$Ga_2O_3$**

| | XX | YY | ZZ | XY | YZ | ZX |
|---|---|---|---|---|---|---|
| x | 0 | 0 | 0 | 0 | 0 | -1.11722 |
| y | 0 | 0 | 0 | 0 | -0.32893 | 0 |
| z | -0.1856 | 0.49575 | -0.8364 | 0 | 0 | 0 |

**Table S4 | Elastic constant ($C$) for O-poalr ε-$Ga_2O_3$**

| | XX | YY | ZZ | XY | YZ | ZX |
|---|---|---|---|---|---|---|
| XX | 285.3177 | 142.4481 | 114.6891 | 0.0368 | 0.0072 | 0.1142 |
| YY | 142.4481 | 251.5959 | 115.8591 | 0.0511 | 0.0062 | 0.1364 |
| ZZ | 114.6891 | 115.8591 | 256.6291 | 0.052 | 0.0055 | 0.1339 |
| XY | 0.0368 | 0.0511 | 0.052 | 64.7815 | 0.0219 | 0.0013 |
| YZ | 0.0072 | 0.0062 | 0.0055 | 0.0219 | 39.2518 | -0.0156 |
| ZX | 0.1142 | 0.1364 | 0.1339 | 0.0013 | -0.0156 | 85.1148 |

piezoelectric strain tensor ($d_{33}$) for O-poalr ε-$Ga_2O_3$:

$$d_{33} = \sum_{j=1}^{6} e_{3j}\,(C^{-1})_{j3}$$

$$d_{33} = e_{31}(C^{-1})_{13} + e_{32}(C^{-1})_{23} + e_{33}(C^{-1})_{33}$$

$$d_{33} = -4.976\ pm/V$$